\documentclass[aps,prb,amsmath,amssymb,footinbib,showpacs,twocolumn,superscriptaddress,longbibliography]{revtex4-2}
\usepackage{amsmath}
\usepackage{amssymb}
\usepackage{amsthm}
\usepackage{balance}
\usepackage{setspace}
\usepackage{graphicx}
\allowdisplaybreaks[4]
\usepackage{braket}
\usepackage{mathrsfs}
\usepackage{float}
\usepackage[colorlinks = true,linkcolor = red,urlcolor  = blue,citecolor = blue,anchorcolor = blue]{hyperref}
\usepackage[utf8]{inputenc}
\usepackage[english]{babel}
\usepackage{commath}
\usepackage{bm}
\usepackage[caption=false]{subfig}

\begin{document}
\title{Intrinsic anomalous thermal hall effect as a signature of quantum metric in $d$-wave altermagnets}

\author{Rishi G. Gopalakrishnan}
\affiliation{Department of Physics and Astronomy, Clemson University, Clemson, SC 29634, USA}
\author{Srimayi Korrapati}
\affiliation{Department of Physics and Astronomy, Clemson University, Clemson, SC 29634, USA}
\author{Sumanta Tewari}
\affiliation{Department of Physics and Astronomy, Clemson University, Clemson, SC 29634, USA}
\begin{abstract}

    We investigate the intrinsic anomalous thermal Hall effect in $d$-wave altermagnets, where a transverse heat current is generated by a longitudinal temperature gradient in the absence of a magnetic field, with the leading response proportional to $(\nabla T)^3$. In these systems, the intrinsic Berry curvature-driven linear and thermal quantum-metric-driven second-order anomalous thermal Hall currents vanish as a consequence of crystalline symmetry. We show that the first nonvanishing contribution arises at third order in the temperature gradient and is governed by a \textit{nonlinear} thermal Berry-connection polarizability, a quantity introduced in this work. Our analysis reveals a distinctive angular dependence of the anomalous thermal Hall conductance as the applied thermal gradient is rotated with respect to the crystal axes. We also find characteristic temperature and chemical-potential dependences that can be tested experimentally. These results identify unique quantum geometry-induced thermal responses and establish altermagnets as a promising platform for exploring intrinsic (i.e., scattering-time-independent) geometric transport phenomena.
\end{abstract}

\maketitle

\section{Introduction}
Altermagnets are a recently identified class of collinear magnets that combine key features of ferromagnets and antiferromagnets while maintaining zero net magnetization \cite{hayami2019momentum, vsmejkal2022beyond, vsmejkal2022emerging, mazin2021prediction, bai2024altermagnetism}. Although altermagnets have vanishing net magnetization, they can nevertheless support anomalous Hall responses \cite{feng2022anomalous, ghorashi2024altermagnetic, fernandes2024topological}. In contrast to conventional antiferromagnets, whose magnetic sublattices are typically related by translation or inversion, altermagnets are characterized by sublattices related by a spin-group element. In the presence of spin-orbit coupling (SOC), the spin-group element becomes the antiunitary magnetic symmetry $\hat{\mathcal{C}}_n\hat{\mathcal{T}}$ \cite{vsmejkal2022beyond, liu2022spin, bhowal2024ferroically, tamang2025altermagnetism, jungwirth2026symmetry}. This symmetry structure gives rise to distinctive transport signatures despite the absence of net magnetization and has made altermagnets a promising platform for spintronics, unconventional superconductivity, and quantum information applications \cite{ouassou2023dc, bai2022observation, bai2024altermagnetism, gonzalez2023spontaneous, thomasaltermagnets, korrapati2025electric, farajollahpour2025light, chakraborti2025zeeman, yang2025observation, hussain2025exploring, korrapati2025approximate, fukaya2025superconducting, ma2024altermagnetic}. \\

Anomalous Hall responses have long served as a sensitive probe of nontrivial band structure-driven geometric effect in topological materials \cite{karplus1954hall, klitzing1980new, liu2016quantum, das2021topological, barman2025intrinsicnonlinearplanarthermal}. In particular, the intrinsic linear order anomalous Hall effect, defined as the scattering time ($\tau$) independent transverse response proportional to the applied epectric field $\bm{E}$, can arise in the absence of an external magnetic field and is governed by the integral of the Berry curvature of the occupied bands, also known as the Chern number \cite{nagaosa2010anomalous, vsmejkal2022anomalous}. It has recently been shown that the Berry curvature dipole (BCD) and quadrupole (BCQ), defined as the first and second moments of the Berry curvature over the occupied states, can generate anomalous Hall responses at higher order in the applied electric field, including contributions with different scattering-time dependence \cite{sodemann2015quantum, you2018berry, sankar2024experimental, zhang2023higher, lai2021third}. Alternatively, an external electric field or temperature gradient can induce higher order corrections to the Berry curvature itself, giving rise to scattering time independent anomalous Hall and thermal Hall responses at higher order in $\bm E$ and $\bm \nabla T$ \cite{wang2021intrinsic, huang2023intrinsic, zhou2022fundamental, varshney2023intrinsic}. In this work, we focus on these scattering time independent contributions, which we refer to as the intrinsic anomalous Hall and intrinsic anomalous thermal Hall responses. \\

In the $d$-wave altermagnet considered in this work, the Brillouin-zone integral of the Berry curvature over the occupied states vanishes as a consequence of the $\hat{\mathcal{C}}_{4z}\hat{\mathcal{T}}$ symmetry, and therefore the linear intrinsic anomalous Hall response is absent. This naturally motivates the study of higher-order anomalous Hall responses as probes of the underlying geometric structure in $d$-wave altermagnets. Recent work has shown that an external electric field can induce corrections to the Berry curvature, thereby giving rise to intrinsic anomalous Hall responses beyond linear order that can, in principle, be accessed experimentally \cite{sankar2024experimental, yu2025quantum, ma2019observation}. In particular, the second-order intrinsic anomalous Hall current is governed by the first-order electric-field-induced correction to the Berry curvature, which can be expressed in terms of the quantum metric dipole (QMD) \cite{PhysRevLett.133.106701}. For the present system, however, the $\hat{C}_{2z}$ spin-group symmetry, consisting of a $\pi$ rotation about the $z$ axis in both real and spin space, forbids even the second-order intrinsic Hall response. As a result, the leading intrinsic anomalous Hall response in this system appears at third order in the applied electric field \cite{korrapati2025electric, korrapati2025approximate}. \\

Analogously, thermal transport provides an additional route to probe quantum geometric effects through thermal Hall and Nernst responses \cite{katsura2010theory, zhang2024thermal, nandy2019planar, sharma2019transverse, behnia2016nernst}. In the $d$-wave altermagnet considered here, the intrinsic linear anomalous thermal Hall response is forbidden by the $\hat{\mathcal{C}}_{4z}\hat{\mathcal{T}}$ symmetry, just as the linear intrinsic anomalous Hall response vanishes in the corresponding charge transport problem. This naturally motivates the study of higher-order thermal Hall effects as probes of the underlying quantum geometry of the Bloch bands. It is by now well established that a temperature gradient can induce corrections to quantum-geometric quantities of Bloch bands \cite{PhysRevB.111.165424, li2024intrinsic}, in close analogy with the corrections generated by an external electric field \cite{gao2014field, gao2015geometrical, liu2025quantum}. Recent work has further proposed a nonlinear planar thermal Hall effect in the presence of an in-plane magnetic field arising from such temperature gradient-induced quantum geometric corrections \cite{barman2025intrinsicnonlinearplanarthermal}.\\

Motivated by these developments, we investigate the intrinsic nonlinear anomalous thermal Hall effect (NATHE), namely a transverse heat current generated in zero magnetic field by a longitudinal temperature gradient through higher-order Berry curvature corrections. In analogy with the anomalous Hall effect, the nonlinear thermal Hall response arises from quantum geometric corrections induced by the thermal gradient. In particular, it was shown that the second-order anomalous thermal Hall current is governed by the Brillouin zone integral of the thermal Berry connection polarizability dipole. For the $d$-wave altermagnet, however, this contribution vanishes because of the $\hat{\mathcal{C}}_{2z}$ symmetry. Consequently, the leading intrinsic anomalous thermal Hall response appears at third order in the temperature gradient, i.e., it is proportional to $(\nabla T)^3$.\\

In this work, we develop a formalism for the third-order intrinsic (i.e., scattering-time-independent) NATHE in altermagnets. Previous studies showed that the first-order correction to the Berry curvature is governed by the thermal Berry connection polarizability (TBCP) tensor [Eq.~\eqref{eq:TBCP}] \cite{li2024intrinsic}, which is the thermal analog of the Berry connection polarizability (BCP) \cite{gao2014field, gao2015geometrical, liu2025quantum}. Here, we show that the second-order correction to the Berry curvature induced by an applied thermal gradient, obtained within the Schrieffer-Wolff perturbative framework, is controlled by a quantity that we define as the nonlinear thermal Berry connection polarizability (NTBCP) tensor [Eq.~\eqref{eq:NTBCP}]. The NTBCP tensor diverges at the band-touching points, and consequently, the associated kernel displayed in Fig.~\ref{fig.modbc} is also singular at those points. Despite this singular behavior at the level of the band resolved kernel, the total third-order thermal current [Eq.~\eqref{jQT}] and the associated response tensors [Eq.~\eqref{transportcoeffeqn}] remain finite. This is because the Brillouin-zone integral over the occupied states [Eq.~\eqref{eq:eta}] contains compensating contributions from the two bands, as reflected in Figs.~\eqref{fig.etavstheta}, \eqref{fig.etavsT}, and \eqref{fig.etavsmu}. Our results therefore identify a distinct quantum-geometric mechanism for nonlinear thermal Hall transport in $d$-wave altermagnets and point to experimentally accessible signatures of the underlying non-trivial quantum geometry.\\ 

The remainder of this paper is organized as follows: In Sec.~\ref{sec: theo}, we derive the energy and Berry-curvature corrections and obtain the third-order anomalous thermal current and the corresponding thermal conductivity tensor for a general two-band Hamiltonian. In Sec.~\ref{ITH}, we introduce the $d$-wave altermagnet Hamiltonian and describe its symmetries. In Sec.~\ref{Results} we apply the formalism in Sec.~\ref{sec: theo} to the $d$-wave altermagnet system and present the resulting thermal Hall response, together with its characteristic experimental signatures. Finally, we summarize our results in Sec.~\ref{sec:con}.

\section{Theoretical Background}
\label{sec: theo}
In this section, we outline the general formalism for deriving the nonlinear thermal Berry connection polarizability tensor (NTBCP) with the help of the the Schrieffer-Wolff perturbation framkework, and analyze the intrinsic third-order anomalous thermal Hall response it generates.
\subsection{Nonlinear thermal Berry connection polarizability tensor}
The total Hamiltonian, in the presence of an applied Thermal gradient is given by $\hat{H} = \hat{H}_0 + \hat{H}'$ \cite{barman2025intrinsicnonlinearplanarthermal, PhysRevB.111.165424}, where $H_0$ is the unperturbed Hamiltonian, and
\begin{eqnarray}
    \hat{H}' &=&  -\frac{1}{2}E_T^a\{\hat{H}_0, \hat{r}^a\}, \label{HpertT} 
\end{eqnarray}
where the notation $\{,\}$ represents the anti-commutation operation, $\hat{r}$ is the position operator, and $\bm{E_T} = -\frac{\bm{\nabla}T}{T}$ represents the thermal gradient. We also use the Einstein's summation convention for the vector indices of the operator throughout the paper. We first expand the Berry curvature and energy up to the second power in the thermal gradient as
\begin{eqnarray}
    \tilde\epsilon_{n,\mathbf{k}} &=&  \epsilon_{n,\mathbf k}^{(0)} + \epsilon_{n,\mathbf k}^{(1)} + \epsilon_{n,\mathbf k}^{(2)} ,\label{Ecorr} \\
    \tilde{\mathbf{\Omega}}_{n,\mathbf k} &=&  \mathbf \Omega_{n,\mathbf k}^{(0)} + \mathbf \Omega_{n,\mathbf k}^{(1)} + \mathbf \Omega_{n,\mathbf k}^{(2)},\label{BCcorr}
\end{eqnarray}
where $\epsilon_{n,\mathbf k}^{(m)}$ and $\Omega_{n,\mathbf k}^{(m)}$ represents the $m^{th}$ order correction in thermal gradient ($\bm{\nabla} T$) to the energy and Berry curvature respectively. To compute these corrections perturbatively, we use the Schrieffer-Wolff perturbation scheme \cite{PhysRevLett.133.106701}. The energy and Berry curvature corrections up to the second order in thermal gradient ($\bm{\nabla} T$) for a two-band system are given below (see appendix for calculations): 
\begin{eqnarray}
    \epsilon_{n,\mathbf k}^{(1)} &=& 0, \label{epsilon1} \\
    \epsilon_{n,\mathbf k}^{(2)} &=& \frac{1}{2}E^a_TE^b_TF^{ab}_{n,T}, \label{epsilon2} \\
    \left(\Omega_n^{(1)}\right)^c &=& -E^d_T\epsilon^{abc}\partial_aG^{bd}_{n, T}, \label{Omega1} \\
    \left(\Omega_n^{(2)}\right)^c &=& 2E^a_TE^b_TF^{ab}_{n,T}\frac{\big(\Omega^{(0)}_n\big)^{c}}{\epsilon_{n\bar{n}}}, \label{Omega2}
\end{eqnarray}
where $\Omega^a_n$ is the unperturbed Berry curvature, $\epsilon_{n\bar{n}} = \epsilon_{n,\mathbf k}^{(0)} - \epsilon_{\bar{n},\mathbf k}^{(0)}$, $G^{ab}_{n, T}$ is the Thermal Berry Connection Polarizability (TBCP) tensor \cite{PhysRevB.111.165424, li2024intrinsic}, and $F^{ab}_{n,T}$ is the Nonlinear Thermal Berry Connection Polarizability (NTBCP) tensor, defined as,
\begin{eqnarray}
    G^{ab}_{n,T} &=& -\frac{\epsilon_{n}^{(0)} + \epsilon_{\bar{n}}^{(0)}}{2\epsilon_{n\bar{n}}}(A^a_{n\bar{n}}A^b_{\bar{n}n} + A^b_{n\bar{n}}A^a_{\bar{n}n}), \label{eq:TBCP} \\
    F^{ab}_{n,T} &=& \frac{(\epsilon_{n}^{(0)} + \epsilon_{\bar{n}}^{(0)})}{2}G^{ab}_{n,T}, \label{eq:NTBCP} 
\end{eqnarray}
where $A^a_{n\bar{n}} = \bra{n}r^a\ket{\bar{n}}$ is the inter-band Berry connection, and $\ket{n}$ is the unperturbed cell-periodic Bloch eigenstate. We shall now apply these results to compute the third order intrinsic anomalous thermal Hall currents in two dimensional $d_{x^2-y^2}$-wave altermagnetic systems. 

\subsection{Third order intrinsic anomalous thermal Hall effect}
In this paper, we focus on the transverse thermal currents in Altermagnets (see Eq.\eqref{Hamil}) in the absence of an applied magnetic field. As a result of the symmetries present in the altermagnet system (see discussion on symmetries in Sec.~\ref{ITH}), the linear and quadratic order transverse thermal currents vanish. Therefore, in this work, we restrict our analysis exclusively to the third order thermal Hall response in the applied thermal gradient $\bm{\nabla}T$ ($\propto (\nabla T)^3$). The total thermal current $\bm{j}^Q_{tot}$, accounting for both the intrinsic and non-intrinsic, obtains contributions from three different sources and can be written as $\bm{j}^Q_{tot} = \bm{j}^Q_{v} + \bm{j}^Q_{E} + \bm{j}^Q_{T}$. Here, the first term $\bm{j}^Q_{v}$ is the contribution to the thermal current arising from the conventional velocity $\bm{v}_k$ of the carriers, the second term $\bm{j}^Q_{E}$ is the anomalous thermal current driven by the non-trivial Berry curvature $\tilde{\bm{\Omega}}_{\bm{k}}$ in the presence of an electric field $\bm E$. Finally, the third term $\bm j^Q_T$ is the anomalous thermal Hall current driven by $\tilde{\bm{\Omega}}_{\bm{k}}$ in the presence of a thermal gradient $\bm \nabla T$ and it is the primary focus of this paper. The expression for this current is given by \cite{PhysRevB.105.125131, PhysRevResearch.2.032066, PhysRevB.111.165424}
\begin{eqnarray}
    \bm j^Q_T &=& -\frac{k_B^2T}{\hbar}\bm{\nabla}T\times\int_{\bm{k}}\tilde{\bm{\Omega}}_{n,\bm{k}}[\beta^2(\tilde{\epsilon}_{\bm{n,k}}-\mu)^2\tilde{f}^n_0 \nonumber \\
    &&+\frac{\pi^2}{3} - \ln^2(1-\tilde{f}^n_0) - 2Li_2(1-\tilde{f}^n_0)], \label{jQT}
\end{eqnarray}
where $\int_{\bm{k}}\equiv\int[d\bm{k}]\sum_n$, $\beta = \frac{1}{k_BT}$, $Li_2(x)$ is the dilogarithm function, and the equilibrium Fermi-Dirac distribution function is expanded to second order in the thermal gradient as
\begin{equation}
\tilde{f}_0^n\equiv f^n_{0}(\tilde{\epsilon}_{n,\bm{k}})=f^n_{0}(\epsilon_{n,\bm{k}})+\epsilon^{(2)}_{n,\bm{k}}\partial_{\epsilon}f^n_{0}(\epsilon_{n,\bm{k}}).    
\end{equation}
By introducing a dummy variable $\varepsilon$ and integrating by parts, we reduce Eq.~\eqref{jQT} to the more concise form \cite{yokoyama2011transverse, qin2011energy}
\begin{equation}\label{eq:current_compact}
    \bm j^Q_T = \frac{1}{\hbar}\frac{\bm{\nabla} T}{T}\times\int_{\bm{k}} \tilde{\bm{\Omega}}_{n,{\bm{k}}}\int_{- \infty}^{\infty}d\epsilon g({\epsilon})\Theta(\epsilon-\tilde{\epsilon}_{n,{\bm{k}}}),
\end{equation}
where 
\begin{equation}
    g(\epsilon)= (\epsilon-\mu)^2 \partial_{\epsilon}f_0
\end{equation}
Expanding the step function to the second order in the thermal gradient, we have
\begin{equation}\label{eq:step_expanded}
    \Theta(\epsilon-\tilde{\epsilon}_{n,{\bm{k}}})=\Theta(\epsilon-\epsilon^{(0)}_{n,{\bm{k}}})-\epsilon^{(2)}_{n,{\bm{k}}}\delta(\epsilon-\epsilon^{(0)}_{n,{\bm{k}}}).
\end{equation}
Recalling that $\tilde{\bm{\Omega}}_{n,\bm{k}} = \bm{\Omega}^{(0)}_{n,\bm{k}} + \bm{\Omega}^{(1)}_{n,\bm{k}} + \bm{\Omega}^{(2)}_{n,\bm{k}}$ contains contributions of zeroth, first, and second order in the thermal gradient, respectively, we analyze the different terms entering the anomalous thermal Hall current in Eq.~(\ref{eq:current_compact}). Noting that the prefactor in Eq.~(\ref{eq:current_compact}) carries one power of the thermal gradient, the first- and second-order intrinsic contributions—arising from $\bm{\Omega}^{(0)}_{n,\bm{k}}$ and $\bm{\Omega}^{(1)}_{n,\bm{k}}$ combined with the first term in Eq.~(\ref{eq:step_expanded}), respectively—vanish due to $\hat{\mathcal{C}}_{4z}\hat{\mathcal{T}}$ symmetry. The leading nonvanishing contribution arises at third order in the thermal gradient, originating from the combination of $\bm{\Omega}^{(2)}_{n,\bm{k}}$ with the first term in Eq.~(\ref{eq:step_expanded}), and $\bm{\Omega}^{(0)}_{n,\bm{k}}$ with the second term ($\propto\epsilon^{(2)}_{n,{\bm{k}}}$). The resulting third order current can be written as 
\begin{eqnarray}
    \left(j^{Q(3)}_{T}\right)_a = \eta_{abcd}  (\nabla T)^b(\nabla T) ^c(\nabla T)^d.\label{transportcoeffeqn}
\end{eqnarray}
where the third order anomalous thermal Hall conductivity tensor components are given by
\begin{equation} \label{eq:eta}
    \eta_{abcd}=-\frac{1}{\hbar T^3}\epsilon_{ab}\int_{\bm{k}} F^{cd}_{n,T}\Omega^{z}_n\bigg(\frac{g(\epsilon^{(0)}_{n,\bm{k}})}{2}-\frac{2}{\epsilon_{n\bar{n}}}\int_{\epsilon^{(0)}_{n,\bm{k}}}^{\infty}d\epsilon g({\epsilon})\bigg).
\end{equation}
We emphasize that the conductivity is independent of scattering time $\tau$ and the corresponding response is intrinsic in nature. Among the 16 possible components of the conductivity tensor, the antisymmetry of $\epsilon_{ab}$ immediately enforces
\begin{equation}\label{eq:eta_vanishing}
    \eta_{aacd}=0,
\end{equation}
leaving eight components identically zero. The remaining eight components satisfy $\eta_{abcd}=-\eta_{bacd}$. Moreover, since $G^{cd}_T$, and consequently $F^{cd}_T$, is symmetric under the interchange of its indices (see Eqs.~(\ref{eq:TBCP})–(\ref{eq:NTBCP})), we obtain $\eta_{abcd} = \eta_{abdc}$, thereby reducing the number of independent components to at most four. We note that the current $\bm j^Q_T$ (Eq.\eqref{jQT}) is always orthogonal to the applied thermal gradient $\bm \nabla T$. Upon symmetrization over the last three indices, $\eta_{abcd}$ becomes antisymmetric under exchange of the first index with any of the remaining indices, thereby characterizing it as a genuine thermal Hall conductivity.
Let us now apply these results to the case of altermagnets. 
\section{Model Hamiltonian: $d$-wave Altermagnet with rashba SOC}
\label{ITH}
In this section, we apply the results derived in the previous section to the case of 2D $d$-wave altermagents. We use the following model hamiltonian for $d$-wave altermagnets:
\begin{eqnarray}
    \mathcal H(\bm{k}) = &-&2t(\cos k_x  + \cos k_y) + 2\lambda(\sin k_y \sigma_x \nonumber \\
    &-& \sin k_x \sigma_y) + 2\Delta_d(\cos k_x - \cos k_y)\sigma_z, \label{Hamil}
\end{eqnarray}

\begin{figure}[t]
    \hspace*{-0.3 cm}
    \subfloat{%
        \includegraphics[scale=0.5]{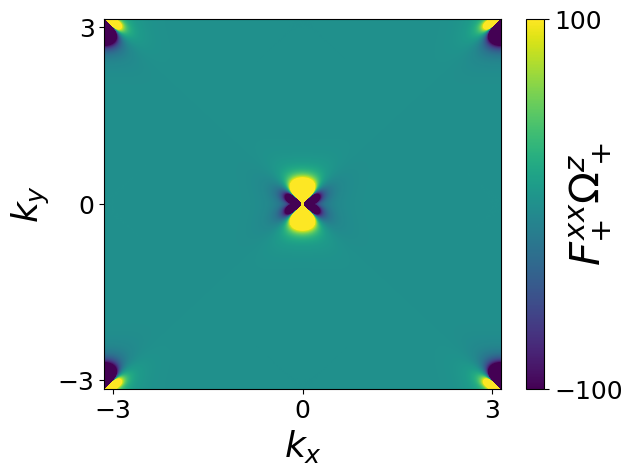}%
    }
    \hspace{0.5 cm}
    \subfloat{%
        \includegraphics[scale=0.5]{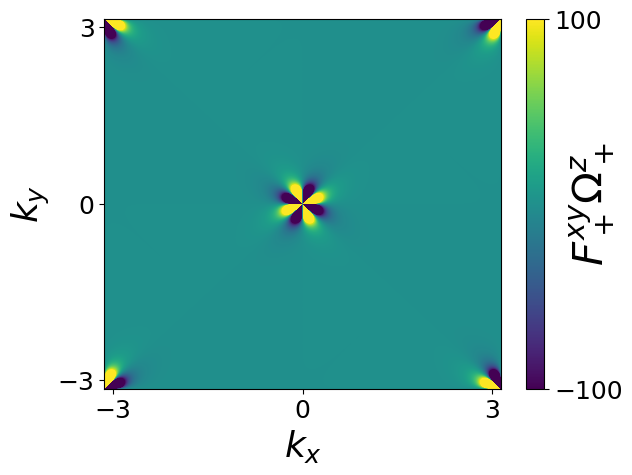}%
    }
    \caption{Variation of the factors $F^{xx}_+\Omega^z_+$ (top panel), $F^{xy}_+\Omega^z_+$ (bottom panel) in Eq.~\eqref{eq:eta} over the $k_x-k_y$ plane. We note that $F^{xx}_{T,=}\Omega^{z}$ is even in $k_x$ and $k_y$ which makes the BZ integral of this quantity finite, resulting in $\eta_{abcc}\neq0$. On the other hand, $F^{xy}_{T,+}\Omega^{z}_+~$ is odd in both $k_x$ and $k_y$ and accordingly integrates to zero, resulting in vanishing  $\eta_{abxy}$ and $\eta_{abyx}$. 
    These factors are concentrated around the the band crossings at $\Gamma\equiv(0,0)$ and $M\equiv(\pi,\pi)$ points where they diverge as expected from the energy difference in the denominators (see Eqs.~(\ref{Omegaij}-\ref{Fij}). The parameters used here are $t=1eV$, $\lambda = 0.3t$, and $\Delta_d = 0.5t$.}
    \label{fig.modbc}
\end{figure}

\noindent where $t$ is the nearest neighbor hopping parameter, $\lambda$ is the Rashba spin orbit coupling strength, and $\Delta_d$ is the altermagnet parameter. For a two-band model, written in the form $\mathcal{H}(\bm{k}) = d_0 + \bm{d}\cdot\bm{\sigma}$, where $d_0$ denotes the kinetic contribution to the Hamiltonian, and $\bm{\sigma}$ is the vector of Pauli matrices. the Berry curvature, TBCP (see Eq.\eqref{eq:TBCP}), and NTBCP (see Eq.\eqref{eq:NTBCP}) are given by the following expressions \cite{graf2021berry}:
\begin{eqnarray}
   \Omega_{\pm}^{ij} &=& \mp\frac{1}{2}\hat{\bm{d}}\cdot(\partial_i\hat{\bm{d}}\times\partial_j\hat{\bm{d}}), \label{Omegaij} \\
   G^{ij}_{T,\pm} &=& \mp\frac{d_0}{8|\bm{d}|}\partial_i\hat{\bm{d}}\cdot \partial_j\hat{\bm{d}}, \label{GTij} \\
   F^{ij}_{T,\pm} &=& \mp\frac{d_{0}^2}{8|\bm{d}|}\partial_i\hat{\bm{d}}\cdot \partial_j\hat{\bm{d}}, \label{Fij}
\end{eqnarray}
where $\hat{\bm{d}} = \bm{d}/|\bm{d}|$, $\Omega^z_{\pm} = (\Omega^{xy}_{\pm} - \Omega^{yx}_{\pm}) /2= \Omega^{xy}_{\pm}$, with $\pm$ labeling the conduction/valence bands. Using the low-energy Hamiltonian expanded near the $\Gamma$-point,
\begin{equation}\label{eq:ham_lin}
    \mathcal{H}(\bm{k})_{\Gamma} = -4t+tk^2+2\lambda k_y\sigma_x-2\lambda k_x \sigma_y + \Delta_{d}(k_y^2-k_x^2)\sigma_z,
\end{equation}
we analytically evaluate the components of the NTBCP tensor for the valence band as
\begin{equation}
    \begin{split}               F^{xx}_{T,-}=&\frac{4\lambda^2d_0^2}{8d_{\bm k}^5}(\Delta_{d}^2k_x^4 + 6\Delta_{d}^2k_x^2k_y^2 + \Delta_{d}^2k_y^4 +4 k_y^2\lambda^2)\\     F^{yy}_{T,-}=&\frac{4\lambda^2d_0^2}{8d_{\bm k}^5}(\Delta_{d}^2k_x^4 + 6\Delta_{d}^2k_x^2k_y^2 + \Delta_{d}^2k_y^4 + 4k_x^2\lambda^2)\\
     F^{xy}_{T,-}=&\frac{-2k_xk_y\lambda^2 d_0^2}{d_{\bm k}^5}(\Delta_{d}^2k_x^2 + \Delta_{d}^2k_y^2 + \lambda^2)
    \end{split}
\end{equation}
and the Berry curvature as 
\begin{equation}
    \Omega^{z}_{-}=\frac{2\lambda^2\Delta_d(k_x^2-k_y^2)}{|\bm{d}|^{3}}
\end{equation}.
We now discuss the symmetries that constrain the allowed responses in the altermagnet system. For the specialized case of a $d$-wave altermagnet with SOC in Eq.~\eqref{Hamil}, the relevant symmetries include the combined operations $\hat{\mathcal{C}}_{4z}\hat{\mathcal{T}}$ and $\hat{\mathcal{M}}_x\hat{\cal{T}}$. Here, time-reversal symmetry is given by $\hat{\mathcal{T}}=-i\sigma_y \hat{\mathcal{K}}$ (with $\hat{\mathcal{K}}$ denoting complex conjugation), along with fourfold rotational symmetry about the $z$-axis, $\hat{\mathcal{C}}_{4z}=e^{i\frac{\pi}{4}\sigma_z}$, and mirror symmetry about the $x$-axis, $\hat{\mathcal{M}}_x=i\sigma_x$. $(\hat{\mathcal{C}}_{4z}\hat{\mathcal{T}})^2 = \hat{\mathcal{C}}_{2z}$ which is the spin group symmetry comprising of a $\pi$ rotation space about the $z$-axis in both the real and the spin spaces. The $\hat{\mathcal{C}}_{4z}\hat{\mathcal{T}}$  symmetry forces the Brillouin-zone integral of the inherent Berry curvature $\Omega^{(0)}$ to vanish and therefore eliminates the first-order anomalous thermal Hall current in $\bm{\nabla}T$. $\hat{\mathcal{C}}_{2z}$ symmetry similarly forbids the Brillouin-zone integrals of terms of the form $\partial_a G^{bd}_{n,T}$, so that the intrinsic second-order anomalous thermal Hall response in $\bm{\nabla}T$ also vanishes (see supplement in \cite{PhysRevLett.133.106701}). The leading nonlinear thermal Hall-type response is therefore third order.  Combining the symmetries $\hat{\mathcal{M}}_x\hat{\mathcal{T}}$ and $\hat{\mathcal{C}}_{4z}\hat{\mathcal{T}}$, we obtain two additional symmetries $(\hat{\mathcal{M}}_x\hat{\mathcal{T}})(\hat{\mathcal{C}}_{4z}\hat{\mathcal{T}}) = \hat{\mathcal{M}}_{x=y}$ and $(\hat{\mathcal{C}}_{4z}\hat{\mathcal{T}})(\hat{\mathcal{M}}_x\hat{\mathcal{T}}) = \hat{\mathcal{M}}_{x=-y}$, corresponding to a mirror operation in real space followed by a $\pi$ rotation in spin space. As discussed below, these symmetries impose characteristic angular constraints on the measurable transverse response. We now turn to the intrinsic third-order thermal Hall response.

\section{Results}
\label{Results}
In the presence of two mirror symmetries, as is the case here, tensor $F^{cd}_T$ satisfies
\begin{equation}
F^{cc}_
T(k_x,k_y)=F^{cc}_T(-k_x,k_y)=F^{cc}_T(k_x,-k_y),    
\end{equation}
while the off-diagonal components satisfy
\begin{equation}
 F^{cd}_T(k_x,k_y)=-F^{cd}_T(-k_x,k_y)=-F^{cd}_T(k_x,-k_y).   
\end{equation}
Meanwhile, $\Omega^{z}(k_x,k_y)$ is even in both $k_x$ and $k_y$. With $\epsilon^{(0)}_{\pm,\bm{k}}$ being even in $\bm{k}$,  the behavior of the integrand of Eq.~\eqref{eq:eta} under momentum inversion is governed solely by the symmetry properties of the product $F^{xx}_{T}\Omega^z_{T}$. Figure~\eqref{fig.modbc} shows the momentum-space distribution of the factors $F^{xx}_{T,+}\Omega^z_{T,+}$ (top panel) and $F^{xy}_{T,+}\Omega^z_{T,+}$ (bottom panel). Both components exhibit divergences at the $\Gamma-$point and the  $M-$point, reflecting the band crossings at these momenta. We note that $F^{xx}_{T}\Omega^{z}$ is even in $k_x$ and $k_y$ which makes the BZ integral of this quantity finite, resulting in $\eta_{abxx}\neq0$. On the other hand, $F^{xy}_{T}\Omega^{z}~(c\neq d)$ is odd in both $k_x$ and $k_y$ and accordingly integrates to zero, resulting in a vanishing  $\eta_{abcd}$. This results in the maximum number of independent components going from four (as discussed after Eq.~(\ref{eq:eta_vanishing})) to at most two. Further, under $\mathcal{C}_{4}\mathcal{T}$ symmetry, $\Omega^{z}(k_x,k_y)\to-\Omega^{z}(-k_y,k_x)$, and $F^{xx}_T \to F^{yy}_T$ and thereby $\int [d\mathbf{k}] F^{xx}_T\Omega^{z}=-\int [d\mathbf{k}] F^{yy}_T\Omega^{z}$. Accordingly, we do not show $F^{yy}_T\Omega^z$ separately in Figure~\eqref{fig.modbc}, since it is obtained from $F^{xx}_+\Omega^z_+$ by a $\pi/2$ rotation followed by a sign reversal. This further reduces the number of independent components to one and the four surviving nonzero components are related as 
\begin{equation}\label{eq:eta_0}
    \eta_{xyxx}=-\eta_{xyyy}=-\eta_{yxxx}=\eta_{yxyy}=\eta_{0}. 
\end{equation}
With both $F^{cd}_{T}$ and $\Omega^{z}$ changing signs across bands, the denominator $\epsilon_{n\bar{n}}$ in Eq.~(\ref{Omega2}) ensures that, for a given momentum, the second order corrections to Berry curvature are equal in magnitude and opposite in sign for the two bands. 
\begin{figure}[t]
    \hspace*{-0.3 cm}
    \subfloat{%
        \includegraphics[scale=0.5]{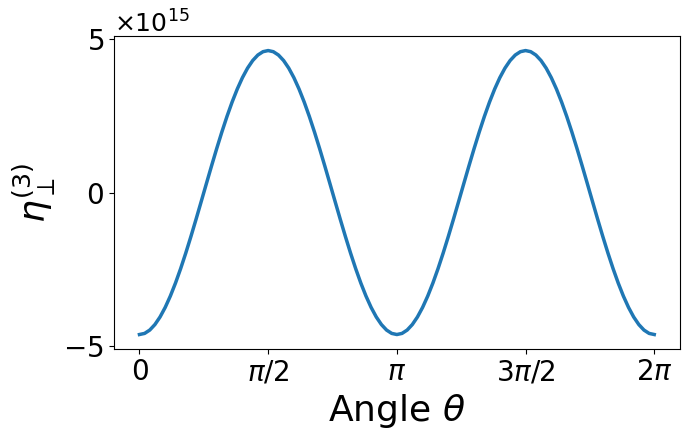}%
    }
    \caption{Variation of the third order thermal Hall conductivity  $\eta_\perp^{(3)}$ in Eq.\eqref{etaperp3} over different values of the polar angle $\theta$ of the applied thermal gradient in the crystal frame. The conductivity $\eta_\perp^{(3)}$ shows a $\cos2\theta$ dependence and vanishes at angle values equal to odd multiples of $\pi/4$. We use the model parameter values $t=1 eV$, $\lambda=0.3t$, $\Delta_d = 0.5t$, chemical potential $\mu = -3.998eV$, and $T = 10K$.}
    \label{fig.etavstheta}
\end{figure}
\begin{figure}[t]
    \hspace*{-0.3 cm}
    \subfloat{%
        \includegraphics[scale=0.5]{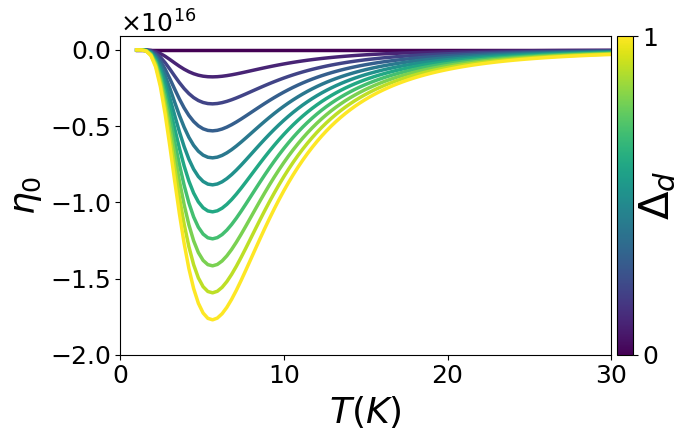}%
    }
    \hspace{0.5 cm}
    \subfloat{%
        \includegraphics[scale=0.5]{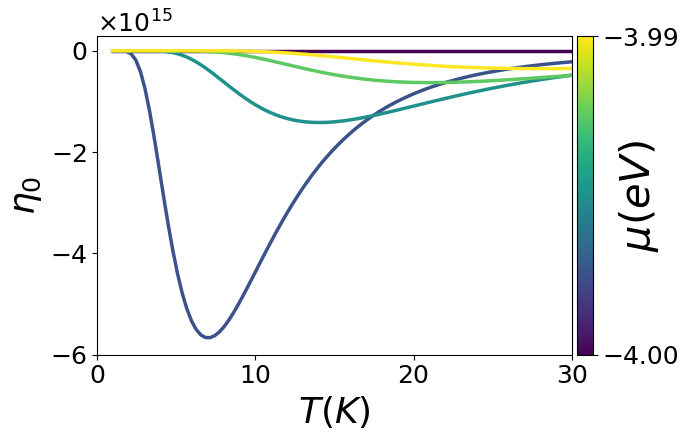}%
    }
    \caption{Variation of the third order thermal Hall conductivity  $\eta_{0}$ as a function of the temperature $T$. Top: Displays the variation of the thermal Hall conductivity for various values of the $d$-wave order parameter $\Delta_d$ with $\mu = -3.998 eV$. As we increase the order parameter, the thermal response becomes stronger. Bottom: Displays the variation of the thermal Hall conductivity for various values of the chemical potential $\mu$ with $\Delta_d = 0.5t$. As the chemical potential moves further from the band crossing point, the magnitude of the peak value of $\eta_{0}$ decreases and the peak also moves to higher temperatures, while keeping the ratio $(\mu-2t)/(k_BT)$ constant. The parameters used here are $t=1eV$, and $\lambda = 0.3t$.}
    \label{fig.etavsT}
\end{figure} 

In this work, we focus on the configuration $\bm \nabla T = \nabla T(\cos\theta, \sin\theta, 0)$, where the thermal gradient is applied at a polar angle $\theta$ in the crystal frame. The third order Hall response in the direction transverse to the applied field is given by  $j^{Q}_\perp(\theta)=\bm{j}^{Q(3)}_T\cdot(\hat{\bm{z}}\times\hat{\bm{E}}_{T})$ and the corresponding thermal-gradient-induced third order thermal Hall conductivity $\eta^{(3)}_{\perp}(\theta)=j^{Q}_\perp(\theta)/(\nabla T)^3$ is explicitly obtained as  
\begin{equation}
\begin{split}
     \eta^{(3)}_{\perp}(\theta)  =\eta_{0}\cos2\theta, \label{etaperp3}
\end{split}
\end{equation}
where we have used Eq.~(\ref{eq:eta_0}). Figure~\eqref{fig.etavstheta} displays the effective transverse thermal conductivity $\eta^{(3)}_{\perp}$, shown in Eq.~\eqref{etaperp3} as a function of the polar angle $\theta$ in the crystal frame. A notable feature is that the effective conductivity $\eta_\perp^{(3)}$ vanishes when the thermal gradient is applied along the diagonal directions, i.e., at odd multiples of $\pi/4$. As a result, the net transverse current also vanishes at these angles. This follows from the $\hat{\mathcal{M}}_{x=-y}$ and $\hat{\mathcal{M}}_{x=y}$ symmetries. For a thermal gradient applied along the diagonal or off-diagonal directions, these symmetry operations reverse the sign of the transverse current, $j_\perp \rightarrow -j_\perp$, while leaving the external configuration unchanged. The transverse current must therefore satisfy $j_\perp=-j_\perp$, which implies $j_\perp=0$. We next analyze some properties of the thermal Hall response tensor $\eta_{abcd}$.

\begin{figure}[t]
    \hspace*{-0.3 cm}
    \subfloat{%
        \includegraphics[scale=0.5]{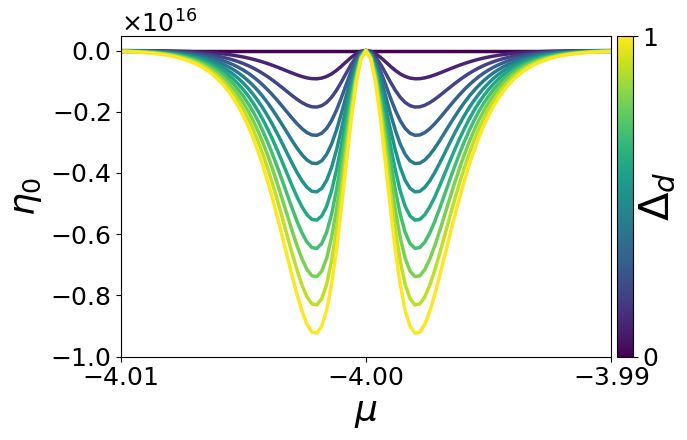}%
    }
    \hspace{0.5 cm}
    \subfloat{%
        \includegraphics[scale=0.5]{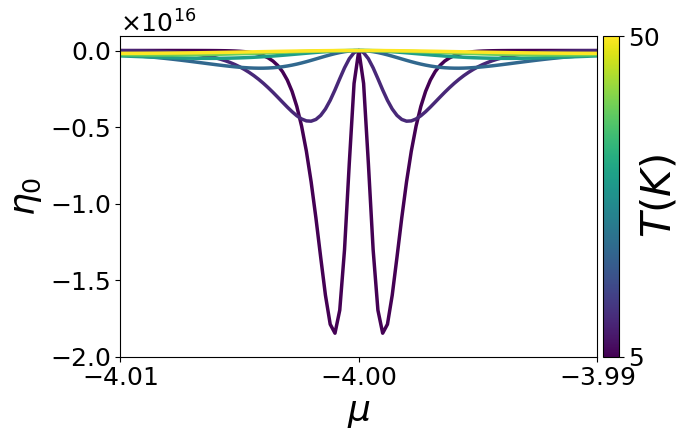}%
    }
    \caption{Variation of the third order thermal Hall conductivity  $\eta_{0}$ as a function of the chemical potential $\mu$. Top: Variation of the thermal Hall conductivity with respect to $\mu$ for various values of the $d$-wave order parameter $\Delta_d$ with $T = 10K$. As we increase the order parameter, the thermal response becomes stronger. Bottom: Variation of the thermal Hall conductivity for various values of the chemical potential $\mu$ with $\Delta_d = 0.5t$. As the temperature increases, the position of the peak of $\eta_{0}$ shifts further from the band crossing point ($\mu = -4.0eV$) and the magnitude of the value of the peak decreases. In both cases, the thermal Hall conductivity vanishes at the band crossing chemical potential and is symmetric about that point. The parameters used here are $t=1eV$, and $\lambda = 0.3t$.}
    \label{fig.etavsmu}
\end{figure}

Figure~\eqref{fig.etavsT} shows the third-order intrinsic thermal Hall conductivity $\eta_{0}$ as a function of temperature for different values of the $d$-wave order parameter $\Delta_d$ (top) and chemical potential $\mu$ (bottom). For fixed $\mu$, the magnitude of $\eta_{0}$ increases from zero with temperature, reaches a maximum, and then decreases. This unimodal behavior can be understood as a consequence of thermal broadening. As the temperature increases, a larger number of states near the band-touching points, where the kernel tensor in Eq.~\eqref{eq:eta} is strongly enhanced, contribute to the response, leading initially to an increase in the thermal Hall tensor. At higher temperatures, however, states from the opposite band are increasingly sampled, and their contributions partially cancel those of the original band, causing the response to decrease. Increasing $\Delta_d$ enhances the overall magnitude of the response, while leaving the peak position essentially unchanged. In contrast, when $\mu$ is moved away from the band-touching energy $E_c=-2t=-4.0\,\mathrm{eV}$, the peak shifts to higher temperatures and its magnitude decreases. At $\mu=E_c$, the net response vanishes because the two bands contribute equally and with opposite sign.\\

This behavior is further illustrated in Fig.~\eqref{fig.etavsmu}, which shows $\eta_{0}$ as a function of chemical potential for several values of $\Delta_d$ (top) and temperature (bottom). The response vanishes at the band-touching energy and is symmetric about it. At fixed temperature, the magnitude of $\eta_{0}$ first increases with $|\mu-E_c|$ and then decreases at larger $|\mu-E_c|$. The nonmonotonic dependence on both temperature and chemical potential reflects a competition between two effects. On the one hand, the kernel tensor in Eq.~\eqref{eq:eta} is strongly enhanced near the band crossing. On the other hand, the states contributing to the transport are sampled over an energy window of order $k_B T$ around the chemical potential. At low temperature, or when $\mu$ lies too far from $E_c$, only a small portion of this window samples the region with large values of the kernel tensor. At higher temperature, contributions from the opposite band become increasingly important and partially cancel the response. The peak therefore occurs when these two tendencies are balanced, giving an approximately constant ratio $k_B T / |\mu-E_c|$.

\section{Conclusion}
\label{sec:con}

In this paper, we investigated the intrinsic nonlinear anomalous thermal Hall effect in the $d$-wave altermagnet described by Eq.~\eqref{Hamil}. Owing to the $\mathcal{C}_{4z}\mathcal{T}$ and $\mathcal{C}_{2z}$  symmetries, the intrinsic anomalous thermal Hall currents vanish at linear and quadratic order in $\bm{\nabla}T$. The leading intrinsic transverse response is therefore third order in $\bm\nabla T$. We derived the corresponding transport coefficients [Eq.~\eqref{eq:eta}] and identified several experimentally accessible signatures of this response in the $d$-wave altermagnet [Figs.~\eqref{fig.etavstheta}, \eqref{fig.etavsT}, and \eqref{fig.etavsmu}].

We first developed the general theory of third-order intrinsic anomalous thermal Hall transport for a generic two-band Hamiltonian. In particular, we derived the Berry curvature and energy correction terms that govern the third-order transverse thermal current [Eqs.~\eqref{Omega1}, \eqref{Omega2}, \eqref{epsilon1}, and \eqref{epsilon2}], together with the corresponding response coefficients [Eq.~\eqref{eq:eta}]. We then specialized this formalism to the $d$-wave altermagnet Hamiltonian in Eq.~\eqref{Hamil}. In this system, the third-order intrinsic anomalous thermal Hall response is controlled by the second-order correction to the Berry curvature induced by the temperature gradient. We showed that this correction is governed by the nonlinear thermal Berry connection polarizability (NTBCP) tensor [Eq.~\eqref{eq:NTBCP}], which becomes singular at the band-touching points. As a result, the band resolved kernel entering Eq.~\eqref{eq:eta} also diverges near these points, as shown in Fig.~\ref{fig.modbc}. Nevertheless, the total thermal current [Eq.~\eqref{jQT}] and the associated anomalous thermal Hall response tensors remain finite because the Brillouin-zone integral contains compensating contributions from the two bands.

The resulting effective third-order thermal Hall coefficient [Eq.~\eqref{etaperp3}] exhibits a characteristic $\cos 2\theta$ angular dependence as the thermal gradient is rotated in the crystal frame [Fig.~\eqref{fig.etavstheta}]. In particular, the transverse thermal current vanishes when the thermal gradient is oriented at odd multiples of $\pi/4$ relative to the crystal axes. We traced this behavior to the mirror symmetries $\hat{\mathcal{M}}_{x=y}$ and $\hat{\mathcal{M}}_{x=-y}$ discussed in Sec.~\ref{ITH}, which force the transverse response to change sign under the corresponding symmetry operations while leaving the applied configuration invariant. These symmetry-enforced nodes provide a clear and experimentally accessible signature of the underlying quantum geometry in the $d$-wave altermagnet.

We further analyzed the third order NATHE response through the dependence of $\eta_{0}$ on chemical potential and temperature [Figs.~\eqref{fig.etavsT} and \eqref{fig.etavsmu}]. As the chemical potential is moved away from the band-touching energy $E_c$, the peak in $\eta_{yxxx}$ shifts to larger values of temperature and decreases in magnitude. Likewise, increasing temperature eventually suppresses the peak response because of the growing cancellation between the two band contributions. Over the parameter range considered here, the peak position is approximately characterized by a constant ratio $k_B T/|\mu-E_c|$. In contrast, increasing the magnitude of the $d$-wave order parameter enhances the overall magnitude of $\eta_{0}$ without qualitatively changing this behavior.

Our results establish the symmetry constraints and characteristic signatures of intrinsic third-order thermal Hall transport in $d$-wave altermagnets. These predictions can be tested experimentally through transverse heat-current measurements under an applied thermal gradient, and they provide a basis for further studies of nonlinear thermal transport in altermagnetic systems.
\section{Acknowledgements}
\label{sec: ack}
We acknowledge support from SC-Quantum, ARO Grant No. W911NF2210247 and ONR Grant No. N00014-23-1-2061.

\appendix
\section{CORRECTION TO THE BERRY CURVATURE AND ENERGY}
\label{appendix: Appendix A}
The total Hamiltonian, in the presence of an applied thermal gradient is given by $\hat{H}_T = \hat{H}_0 +  \hat{H}'$, where $H_0$ is the unperturbed Hamiltonain, and
\begin{eqnarray}
    \hat{H}'_T &=&  -\frac{1}{2}E_T^a\{\hat{H}_0, \hat{r}^a\}, \label{App:HpertT} 
\end{eqnarray}
where the notation $\{,\}$ represents the anti-commutation operation, $\hat{r}$ is the position operator, and $\bm{E_T} = -\frac{\bm{\nabla}T}{T}$ is the thermal gradient. We also use the Einstein's summation convention for the vector indices of the operator throughout the paper. To perturbatively find corrections to the energy ($\epsilon_n$) and Berry curvature ($\Omega_n^c$), we apply the Schrieffer-Wolff transformation. We first separate the Hamiltonian into its diagonal and off-diagonal as,
\begin{eqnarray}
    \hat{H}_0 &=& \sum_n \left(\epsilon_n^{(0)} - \epsilon^{(0)}_nE_T^aA_n^a\right) \ket{n}\bra{n} \label{App:baseHam}\\
    \hat{H}_1 &=& \sum_{\substack{m,n \\ m\neq n}} \left(\frac{1}{2}(\epsilon_n^{(0)} + \epsilon_m^{(0)})E^a_TA^a_{nm}\right)\ket{n}\bra{m} \nonumber \\
    \label{App:pertHam}
\end{eqnarray}
The Schrieffer-Wolff transformation is effectively used to expand an operator $\mathcal{O}$ perturbatively in powers of $\mathcal{S}$ as

\begin{eqnarray}
    \mathcal{O} \rightarrow e^\mathcal{S}\mathcal{O}e^{-\mathcal{S}} = \mathcal{O} + \left[\mathcal{S},\mathcal{O}\right] + \frac{1}{2}\left[\mathcal{S},\left[\mathcal{S},\mathcal{O}\right]\right] + ... \label{App:Opert}
\end{eqnarray}    
This operator $\mathcal{S}$ is chosen such that the Hamiltonian is diagonal upto to the first order in perturbation. Applying the Schrieffer Wolff transformation (see Eq.\eqref{App:Opert}) to the Hamiltonian ($\hat{H}_T'$) operator we get
\begin{eqnarray}
    H' = \hat{H}_T + \left[\mathcal{S},\hat{H}_T\right] + \frac{1}{2}\left[\mathcal{S},\left[\mathcal{S},\hat{H}_T\right]\right] + ... \label{App:Hpert}
\end{eqnarray}
Writing $\hat{H}_T = \hat{H}_0 + \hat{H}_1$ and diagonalizing the Hamiltonian to the first order in the applied thermal gradient, we get
\begin{eqnarray}
    \hat{H}_1 = \left[\hat{H}_0,\mathcal{S}\right] \label{App:Sdefn}
\end{eqnarray}
Substituting Eq.\eqref{App:Sdefn} in Eq.\eqref{App:Hpert} we get
\begin{eqnarray}
    H' = \hat{H}_0 + \frac{1}{2} \left[\mathcal{S},\hat{H}_1\right]  + ... \label{App:Hprime}
\end{eqnarray}
\noindent Finally, solving for S using Eq.\eqref{App:Sdefn}, we obtain
\begin{eqnarray}
    \mathcal{S}_{nn} &=& 0 \label{App:Sdiag} \\
    \mathcal{S}_{nm} &=&  -\frac{\frac{1}{2}(\epsilon_n^{(0)} + \epsilon_m^{(0)})E_T^aA^a_{nm}}{\epsilon^{(0)}_{nm} - E^a_T(\epsilon^{(0)}_nA_n^a - \epsilon^{(0)}_mA_m^a)} \nonumber \\
    &\approx& - \frac{1}{2}\left(\frac{\epsilon_n^{(0)} + \epsilon_m^{(0)}}{\epsilon_{nm}^{(0)}}\right) E^a_TA^a_{nm}  \nonumber \\ 
    &&- \frac{1}{2}\frac{(\epsilon_n^{(0)} + \epsilon_m^{(0)})}{(\epsilon_{nm}^{(0)})^2}E^a_TE^b_T(\epsilon^{(0)}_nA_n^a - \epsilon^{(0)}_mA_m^a)A^b_{nm} \nonumber \\
    && + \space \space ... \label{App:Soffdiag}
\end{eqnarray}  

\noindent We can thus separate $\mathcal{S}_{nm}$ as
\begin{eqnarray}
    \mathcal{S}_{nm} = \sum_{k\in \mathbb{N}}\mathcal{S}_{nm}^{(k)},
\end{eqnarray}
where $\mathcal{S}_{nm}^{(k)}$ is the $k^{th}$ order in perturbation of the operator $\mathcal{S}$ with respect to $\bm{E_T}$. The first and second order results for the expansion of $\mathcal{S}_{nm}$ is given below
\begin{eqnarray}
    \mathcal{S}_{nm}^{(1)} &=&  - \frac{1}{2}\left(\frac{\epsilon_n^{(0)} + \epsilon_m^{(0)}}{\epsilon_{nm}^{(0)}}\right) E^a_TA^a_{nm} \label{App:S1}\\
    \mathcal{S}_{nm}^{(2)} &=&  - \frac{1}{2}\frac{(\epsilon_n^{(0)} + \epsilon_m^{(0)})}{(\epsilon_{nm}^{(0)})^2}E^a_TE^b_T(\epsilon^{(0)}_nA_n^a - \epsilon^{(0)}_mA_m^a)A^b_{nm} \nonumber \\
    \label{App:S2}
\end{eqnarray}

Applying the perturbation scheme in Eq.\eqref{App:Opert} to the Berry connection $A^a_n$, we get
\begin{eqnarray}
    A^{a(1)}_n &=& \braket{n|[\mathcal{S}, \hat{r}^a]|n} \nonumber \\
    &=& \sum_{m\neq n} (\mathcal{S}_{nm}A^a_{mn} - A^a_{nm}\mathcal{S}_{mn}) \nonumber \\
    &=& -\sum_{m\neq n}\left[\frac{1}{2}\frac{E^b_T}{\epsilon_{nm}^{(0)}}(\epsilon_n^{(0)} + \epsilon_m^{(0)})(A^a_{nm}A^b_{mn} + A^b_{nm}A^a_{mn}) \right] \nonumber \\
    &=& E^b_TG^{ba}_{n,T}, \label{App:A1}
\end{eqnarray}
where $G^{ba}_{n,T}$ is the Thermal Berry Connection Polarizability (TBCP) defined as
\begin{eqnarray}
G^{ba}_{n,T} &=& -\sum_{m\neq n}\frac{\epsilon_{n}^{(0)} + \epsilon_{m}^{(0)}}{2\epsilon_{nm}^{(0)}}(A^a_{nm}A^b_{mn} + A^b_{nm}A^a_{mn}) \nonumber \\
\label{App:TBCP}
\end{eqnarray}

Before proceeding with the calculation for the second order correction for to the Berry connection, we first note that the Berry connection $A^a_n$ is a gauge dependent quantity. As a result of this, we obtain unphysical terms in our calculations. To circumvent this issue, we choose a gauge where $\bm{E_T}\cdot \bm{A}^a_n$ is zero.  Under this choice of gauge, the quantity $\mathcal{S}^{(2)}_{nm}$ (see Eq.\eqref{App:S2}) vanishes. Let us now compute the second order correction to the Berry connection. 

To derive the second order correction to the Berry connection, we use the second term in Eq.\eqref{App:Opert} to obtain
\begin{eqnarray}
    &&A_n^{a(2)} = \frac{1}{2}\braket{n|[\mathcal{S}^{(1)},[\mathcal{S}^{(1)},\hat{r}^a]]|n} \nonumber \\
    &=& \frac{1}{2} \sum_{\substack{l,m \\l\neq m\neq n}}(\mathcal{S}^{(1)}_{nl}\mathcal{S}^{(1)}_{lm}A^a_{mn} + A^a_{nl}\mathcal{S}^{(1)}_{lm}\mathcal{S}^{(1)}_{mn} - 2\mathcal{S}^{(1)}_{nl}A^a_{lm}\mathcal{S}^{(1)}_{mn}) + \nonumber \\
    && \frac{1}{2}\sum_{m\neq n}(\mathcal{S}^{(1)}_{nm}\mathcal{S}^{(1)}_{mn}A^a_n + \mathcal{S}^{(1)}_{nm}\mathcal{S}^{(1)}_{mn}A^a_n - 2\mathcal{S}^{(1)}_{nm}\mathcal{S}^{(1)}_{mn}A^a_m) \label{App:A2pre}
\end{eqnarray}
In the two-band model, the expression in Eq.\eqref{App:A2pre} simplifies to
\begin{eqnarray}
    A_n^{a(2)} = \mathcal{S}^{(1)}_{n\bar{n}}\mathcal{S}^{(1)}_{\bar{n}n}(A^a_n - A^a_{\bar{n}}),\label{App:A2pre}
\end{eqnarray}
where $\bar{n}$ is the band that is not $n$ in the two-band model. Substituting Eq.\eqref{App:S1} in Eq.\eqref{App:A2pre}, we get 

\begin{eqnarray}
    A_n^{c(2)} &=& 
 E^a_TE^b_T\left[F^{ab}_{n,T}\left(\frac{A^c_n - A^c_{\bar{n}}}{\epsilon_{n\bar{n}}}\right)\right] \label{App:A2}
\end{eqnarray}
where,
\begin{eqnarray}
    F^{ba}_{n,T} &=& -\sum_{m\neq n}\frac{(\epsilon_{n}^{(0)} + \epsilon_{m}^{(0)})^2}{4\epsilon_{nm}^{(0)}}(A^a_{nm}A^b_{mn} + A^b_{nm}A^a_{mn}) \nonumber \\
    \label{App:NTBCP}
\end{eqnarray}
Thus, the corresponding Berry curvature obtained from Eqs.\eqref{App:A1} and \eqref{App:A2} are
\begin{eqnarray}
    \left(\Omega_n^{(1)}\right)^c &=& E_T^d\epsilon^{abc}\partial_aG^{bd}_{n,T} \label{App:Omega1} \\
      \left(\Omega_n^{(2)}\right)^c &=& E^a_TE^b_T \left[F^{ab}_{n,T}\left(\frac{\Omega^c_n - \Omega^c_{\bar{n}}}{\epsilon_{n\bar{n}}}\right)\right] 
    \label{App:Omega2}
\end{eqnarray}
We now apply this perturbation scheme to find the corrections to the energy $\epsilon_n^{(1)}$ and $\epsilon_n^{(2)}$. We note that $\epsilon_n = \braket{n|H'|n}$ and expanding this quantity in orders of the driving fields ($\bm{E}$ and $\bm{\nabla T}$) we get
\begin{eqnarray}
    \epsilon_n^{(1)} &=& -\epsilon^{(0)}_nE_T^aA_n^a \label{App:epsilon1pre} \\
    \epsilon_n^{(2)} &=& 
     -\frac{1}{2} \sum_{m\neq n}\left[\frac{E^a_TE^b_T}{\epsilon^{(0)}_{mn}}(\epsilon^{(0)}_n+ \epsilon^{(0)}_m)(A^a_{nm}A^b_{mn} + A^b_{nm}A^a_{mn})\right] \nonumber \\
    &=& \frac{1}{2}E^a_TE^b_TF^{ab}_{n,T} \label{App:epsilon2pre}
\end{eqnarray}
Applying the gauge choice $\bm{E_T}\cdot \bm{A}^a_n$ we get
\begin{eqnarray}
    \epsilon_n^{(1)} &=& 0 \label{App:epsilon1} \\
    \epsilon^{(2)}_n &=&  \frac{1}{2}E^a_TE^b_TF^{ab}_{n,T}
    \label{App:epsilon2}
\end{eqnarray}

\bibliography{references-6}
\end{document}